\documentclass{aa}  

\usepackage{graphicx}
\usepackage{txfonts}
\newcommand{\re}{$R_e$}
\newcommand{\Lsig}{$L-\sigma$}
\newcommand{\Lsiga}{$L=L_0\sigma^{\alpha}$}

\newcommand{\Lsigb}{$L=L'_0\sigma^{\beta}$}
\newcommand{\Lsigbtempo}{$L=L'_{0}(t) \sigma^{\beta(t)}$}                   

\newcommand{\MRa}{$R_e$--$M_*$}

\newcommand{\muem}{$\langle\mu\rangle_e$}
\newcommand{\Ie}{$\langle I\rangle_e$}

\newcommand{\IeRe}{$I_e - R_e$}
\newcommand{\IeSig}{$I_e - \sigma$}

\newcommand{\vfilt}{$\rm{V}$}

\newcommand{\ie}{{\em i.e.}}
\newcommand{\eg}{{\em e.g.}}

\begin{document}

   \title{The Fundamental Plane in the hierarchical context}


   \author{M. D'Onofrio\fnmsep\thanks{Corresponding author: Mauro D'Onofrio}
          \inst{1}
          \and
          C. Chiosi\inst{1}
          }

   \institute{Department of Physics and Astronomy, University of Padua,
              Vicolo Osservatorio 3, I35122 Padova (Italy)\\
              \email{mauro.donofrio@unipd.it} \\
              \email{cesare.chiosi@unipd.it}
             }

   \date{Received December 30, 2021; accepted ...}

 
  \abstract
   {The Fundamental Plane (FP) relation and the distribution of early-type galaxies (ETGs) in the FP projections, cannot be easily explained in the hierarchical framework, where galaxies grow up by merging and star formation episodes.}
   { {We want to show here that both the FP and its projections arise naturally from the combination of the Virial Theorem  (VT) and a new time-dependent relation, describing how luminosity and stellar velocity dispersion change during galaxy evolution. This relation has the form of the Faber-Jackson (FJ) relation but a different physical meaning: the new relation is \Lsigbtempo, where its coefficients $L'_0$ and $\beta$ are time-dependent and can vary considerably from object to object, at variance with those obtained from the fit of the \Lsig\ plane. } }
   {We derive the equations of the FP and its projections as a function of { $\beta$} and $L'_0$, by combining the VT and \Lsigbtempo\ law. Then, from their combination we derive the expression of the FP as a function of $\beta$ and the solutions for $\beta$ and $L'_0$.}
   {We demonstrate that the observed properties of ETGs in the FP and its projections can be understood in terms of variations of $\beta$ and $L'_0$. These two parameters encrypt the history of galaxy evolution across the cosmic epochs and determine the future aspect of the FP and its projections. Using the solutions found for $\beta$ and $L'_0$ for each galaxy at the present epoch, we derive the coefficients of the FP (and FJ relation) and show that, the values of the coefficients coming form the fit, obtained in the literature, originate from the average of the single FP coefficients derived for each galaxy. In addition, we show that the variations of $\beta$ naturally explain the curvature observed in the FP projections and the correct position of the Zone of Exclusion.}
   {}


   \maketitle
%

\section{Introduction}\label{sec:1}

The Fundamental Plane (FP) of early-type galaxies (ETGs), \ie\ the correlation observed between the effective radius \re\ (the radius of the circle enclosing half the total luminosity of a galaxy), the mean effective surface brightness \muem\ {(the average surface brightness within \re, here indicated by $\langle I_e \rangle$ or simply $I_e$)} and the central velocity dispersion of stars $\sigma$ (or its average within \re) \citep{DjorgovskiDavis1987,Dressleretal1987}, 
has found in the past numerous astrophysical applications, \eg\ as distance indicator \citep[\eg][]{Donofrioetal1997}, or for the mapping of the peculiar velocity field of galaxies \citep[\eg][]{Willicketal1995,StraussWillick1995}, for predicting the size of lensed objects, and as a diagnostic tool of galaxy evolution \citep[\eg][]{vanDokkumFranx1996,vanDokkumvanderMarel2007,Holdenetal2010}.  

The FP can be written as.

\begin{equation}
    \log(R_e) = a \log\sigma + b \log I_e +c
    \label{eqfp}
\end{equation}
\noindent
where the coefficients $a$, $b$ and $c$ are respectively the slope and the zero-point of the plane. 

Since its discovery, the physical origin of the FP has been always connected with the Virial Theorem (VT), \ie\ with the condition of dynamical relaxation of these stellar systems \citep{Faberetal1987}.
From the VT one expects the values of
the slope $a=2$ and $b=-1$, two values that are a bit different from those obtained from the fit of the observed distribution of ETGs ($a\sim1.2$ and $b\sim-0.8$). In the astronomical literature this difference is known as the problem of the FP tilt \citep[see \eg][among many others]{Faberetal1987,Ciotti1991,Jorgensenetal1996,Cappellarietal2006,Donofrioetal2006,Boltonetal2007}.

The origin of the tilt has been attributed to several physical mechanisms, invoking alternatively (or in mixed proportion): 1) a progressive change of the stellar mass-to-light { ratio $M_*/L$ } of ETGs along the plane \citep[see \eg][]{Faberetal1987,vanDokkumFranx1996,Cappellarietal2006,vanDokkumvanderMarel2007, Holdenetal2010,deGraafetal2021}, 2) the existence of structural and dynamical non-homology in the stellar systems \citep[see \eg][]{PrugnielSimien1997,Busarelloetal1998,Trujilloetal2004,Donofrioetal2008}, 3) the dark matter (DM) content \citep[see \eg][]{Ciottietal1996,Borrielloetal2003,Tortoraetal2009, Taranuetal2015,deGraafetal2021}, {  4) the star formation history and the initial mass function \citep[see \eg][]{RenziniCiotti1993,Chiosietal1998,ChiosiCarraro2002,Allansonetal2009},} 5) the effect of the environment \citep[see \eg][]{Luceyetal1991,deCarvalhoDjorgovski1992,Bernardietal2003, Donofrioetal2008, LaBarberaetal2010, Ibarra-MedelLopez-Cruz2011, Samiretal2016}. 

The small intrinsic scatter around the FP relation ($\approx0.05$ dex in the V-band, \citet{Bernardietal2003}) is also a long debated problem. This scatter is a bit lower than that measured for the Faber-Jackson (FJ) relation \citep{FaberJackson1976} ($\approx0.09$ dex). Some studies claim that the scatter is due to the variation in the formation epoch, others to the dark matter content or to metallicity trends in the stellar populations. Notably, the position of a galaxy above or below the plane seems independent of flattening, anisotropy and isophotal twisting. According to \cite{Faberetal1987}
the deviations from the plane can be associated to variations of the mass-to-light ratio $M/L$, metallicity or age trends in the stellar populations, dynamical and structural properties, and dark matter content and distribution. Several studies confirmed that the variations in the stellar populations are partially responsible for the intrinsic scatter \citep[see \eg,][]{Gregg1992, GuzmanLuceyBower1993}, but also the variation in galaxies age at a given mass has been often invoked to explain the scatter
\citep{Forbesetal1998, Redaetal2005, Magoulasetal2012}. In general, the most accepted idea is that the major part of the scatter is due to variations in $M/L$ ratio \citep{Cappellarietal2006, Boltonetal2008, Augeretal2010, Cappellarietal2013}: some galaxies deviate from the FP for their lower $M/L$ ratio due to younger stellar populations probably formed in recent gaseous mergers.

While many of these explanations for the tilt and the scatter are well suited for a Universe in which galaxies form in single monolithic collapses, there are some problems to accept this solution when we consider the currently preferred hierarchical scenario, in which galaxies assemble their mass in numerous events of merging and bursts of star formation induced by encounters. All these random events are difficult to reconcile with a systematic variation of the galaxy properties along the FP.

Indeed the examination of the FP of ETGs formed through mergers has 
obtained contrasting results. 
\citet{Nipotietal2003}, among the first, explored the effects of dissipation-less merging on the FP using a N-body code. By investigating two extreme cases (equal mass merging and small accretions), they found that the FP is preserved in the case of major mergers, while the accretion of small objects produce a substantial thickening of the plane, in particular when the angular momentum of the system is low. They pointed out that both the FJ and Kormendy \citep{Kormendy1977} relations are not well reproduced by these simulations.

Along the same vein, \cite{robertsonetal06}
showed that gas dissipation provides an important contribution to the tilt of the FP. In their work the dissipation-less
mergers seem to produce galaxies that share the same plane delineated by the virial relation and only when the
gas content is increased (up to $\sim$30\%), the tilt of the FP can be reproduced.
They estimated that $\sim40\div100$ per cent of the tilt is induced by the structural properties of galaxies and not by stellar population effects. This occurs for a trend in the mass ratio $M_{T}/M_*$ (total over stellar) induced by dissipation. 

\cite{Novak2008} also studied the binary mergers of gas-rich disk galaxies obtaining remnants with properties similar to low-mass fast-rotating ellipticals (that are $\sim80\%$ of the ETGs), but was unable to reproduce the high mass nearly spherical non-rotating systems (the remaining $20\%$ of ETGs). He claimed that the only way to reproduce these galaxies, is to deal with realistic mass assembly histories, made of non-regular sequence of mergers with progressively decreasing mass ratios.

Finally, \cite{Taranuetal2015} showed that multiple dry mergers of spiral galaxies can reproduce the central dominant bright ETGs (BCGs) lying along a tight and tilted FP (albeit somewhat less than observed). This support the idea that in high-density environments, gravitation is the only responsible for the formation of the BCGs and it is not necessary to have abundant ($\sim30-40\%$) gas fractions in the progenitors, as suggested by \cite{robertsonetal06} and \cite{Hopkinsetal2008}.

In general, the modern large scale hydro-dynamical simulations, like \eg\ Illustris \citep{Vogelsbergeretal2014}, are able to reproduced several ETGs with the correct velocity dispersions and luminosities. Notably, the simulation correctly predict the tilt of the FP and the evolution of its coefficients with redshift \cite[see][]{Luetal2020}, even if in the FP projections the faint ETGs have
systematically larger radii with respect to observations \citep{Donofrioetal2020}.

We believe that the problem of the FP tilt should not be disjoint from that of reproducing the correct distributions of ETGs in all FP projections. What we need is a theoretical framework, in the context of the hierarchical scenario, able of reproducing the correct FP tilt, together with the observed distributions of galaxies in the FP projections, including both the curvature of the relations and the presence of the Zone of Exclusion (ZoE), the region empty of galaxies that characterizes some of these diagrams. 
All these observational facts need a common and unique explanation.

The aim of this paper is to show that one can achieve such result by adopting an empirical approach, that permits to link the tilt of the FP and all the characteristics of the ETGs distribution in the FP projections in one single framework, within the scheme of galaxy formation predicted by the hierarchical scenario.

The starting point of this approach is to consider separately the role played by gravitation to that more directly linked to luminosity evolution, and to accept that the structural parameters of galaxies are time-dependent. The parameter space that better summarize the role of mass and luminosity as physical time-dependent variables, 
is the \Lsig\ plane. In this plane, the velocity dispersion is a direct proxy of the total mass. Both variables strongly depend on the mass assembly and star formation episodes. Numerical simulations suggest that these variables can either increase or decrease during mergers, encounters and star formation events. Both variables are non-linear function of time. It is this time variability that is absent from our present view of the scaling relations of galaxies. 

Following \citet{Donofrioetal2017} and \citet{Donofrioetal2020}, we argue here that the time evolution of luminosity and velocity dispersion in this plane, can be catch by {the \Lsigbtempo\ relation}, a new relation formally equivalent to the FJ, but with a profoundly different physical interpretation. In this relation $\beta(t)$ and $L'_{0}(t)$ are free time-dependent parameters that can vary considerably from galaxy to galaxy, according to the mass assembly history of each object. The parameter $\beta$ in particular, { defines the direction (not the orientation) of  ETGs evolution}  in every parameter space linked to the FP. {In the following we will drop the time-dependence notation for $L'_0$ and $\beta$, and simply write \Lsigb. }  

We will see in this { study that, accepting the hypothesis of writing the  \Lsig\ law in this time-dependent way, \ie\ adopting the \Lsigb\ relation (an hypothesis confirmed by the data of the Illustris simulations)}, we can obtain a unique explanation for the FP problem and its projections. We will get such result by combining the \Lsigb\ law with the VT, assuming that all variables are time-dependent. 

The paper presents in Sec. \ref{sec:2} the data sample to our disposal for this analysis, in Sec. \ref{sec:3} the mathematical analysis that provides the correct formulation of the FP and its projections, as well as the plots and tables that explain why this approach works so well. Conclusions are drawn in Sec. \ref{sec:4}.

\section{The sample}\label{sec:2}

The observational data for this { study} are extracted from the WINGS and Omega-WINGS databases
\citep{Fasanoetal2006,Varela2009,Cava2009,Valentinuzzi2009,Moretti2014,Donofrio2014,Gullieuszik2015,Morettietal2017,Cariddietal2018,Bivianoetal2017}.
Both surveys have gathered optical and spectroscopic data for many galaxies in nearby clusters (with $0<z<0.07$), measuring several physical parameters: magnitudes, morphological types, effective radii, effective surface brightness, stellar velocity dispersion, redshift, star formation rate, velocity dispersion, S\'ersic index, etc.

These data are used here in different ways according to the plot of interest. 
The reason is that the spectroscopic data-set is a sub-sample of the whole optical sample. For this reason the variables like \eg\ $\sigma$, the stellar mass $M_*$, the star formation rate $SFR$, the redshift, etc. are less numerous than those extracted from the photometric analysis (\eg\ \re, \Ie, $n$, $L_V$, etc.).
Such incompleteness, however, does not affect in any way our results, since for our goal it is sufficient that the data are { numerous enough to show the main trends of the ETG distribution in the FP projections}. Needless to say, that these data are well tested in many previous publications of the WINGS team and are in good agreement with previous literature. The completeness of the data sample is not required here, because we dot not attempt any statistical analysis of the data, nor we fit any distribution. We simply want to show how the different observed distributions of ETGs might find their origin in the combination of the VT and the \Lsigb\ law, according to the different values of $\beta$ and $L'_0$.

The data used are:
1) The velocity dispersion $\sigma$, that are available for $\sim1700$ ETGs. They were already used by \cite{Donofrioetal2008} to infer the properties of the FP. To these data we added the data-set of $\sigma$ derived by \cite{Bettoni2016} for several dwarf ETGs ($\sim20$ objects). The $\sigma$ have been measured within a circular area of 3 arcsec around the center of the galaxy (as in the Sloan Digital Sky Survey)\footnote{The use of different $\sigma$, such as $\sigma_e$ (the velocity dispersion within \re), does not alter in any way the conclusions obtained here. The only effect of such change of variable is to vary systematically all the plot shown here, without affecting the main conclusions.}. 

2) Luminosity, effective radius and effective surface brightness in the V-band of several thousand ETGs. These were derived by \cite{Donofrio2014} with the software GASPHOT \citep{Pignatelli};

3) The distance of the galaxies  derived from the redshifts measured by \cite{Cava2009} and \cite{Morettietal2017}.

4) The stellar masses obtained by \citet{Fritzetal2007}, only for the galaxies of the southern hemisphere. The cross-match between the spectroscopic and optical sample provides here only 480 ETGs with available stellar mass, velocity dispersion, S\'ersic index, effective radius and effective surface brightness.

The error of these parameters is $\simeq20\%$. These are not shown in our plots, because they are much lower than the observed range of variation of the structural parameters and do not affect the whole distribution of ETGs.

In addition to the real data, we have used several artificial data of the Illustris simulation 
\citep[see e.g.][]{Vogelsbergeretal2014,Genel_etal_2014,Nelsonetal2015}. This data-set contains $\sim 2400$ galaxies and it is marked in all plots by yellow dots.
A full description of these data is given in \cite{Cariddietal2018} and \cite{Donofrioetal2020}.
We used the Illustris-1 run with full-physics (with both baryonic and dark matter) having the highest degree of resolution \citep[see Table 1 of][]{Vogelsbergeretal2014}.\footnote{The new IllustrisTNG data are substantially equivalent, being the problem of the larger effective radii of faint galaxies, not completely solved.} We have extract from the Illustris database the \vfilt-band luminosity of the galaxies, the stellar mass, the velocity dispersion, the half-mass radius of the stellar particles, the morphological types, etc.
The projected light and mass profiles of these ETGs have been studied by \cite{Donofrioetal2020}, who computed much robust values for the effective radius \re, effective surface brightness \muem, and best-fit S\'ersic index by constructing the growth curve of each galaxy. 

As mentioned above, the Illustris models are a bit in conflict with the real data in the case of the effective radius $R_e$ { \citep{Donofrioetal2020}.}  However, a careful comparison with the WINGS data showed that they nicely reproduce the bright tail of the \IeRe\ plane and the tail of the \MRa\ relation of the bright ETGs. They fail instead to reproduce the location of dwarf galaxies (masses in the interval $10^8$ to $10^{10} M_\odot$) on both planes. The relative larger radius of these latter galaxies can likely be explained { as due to either still not good treatment of the feedback effects or the presence of galactic winds }  
\citep[see][for more details]{Chiosietal2020}. The velocity dispersion of the Illustris models, on the other hand, are in good agreement with the observed ones, as well as the FJ relation based on them (see below). 
For these reasons we feel safe in using the Illustris models for the purposes of the present study.

   \begin{figure*}
   \centering
   \includegraphics[scale=0.42]{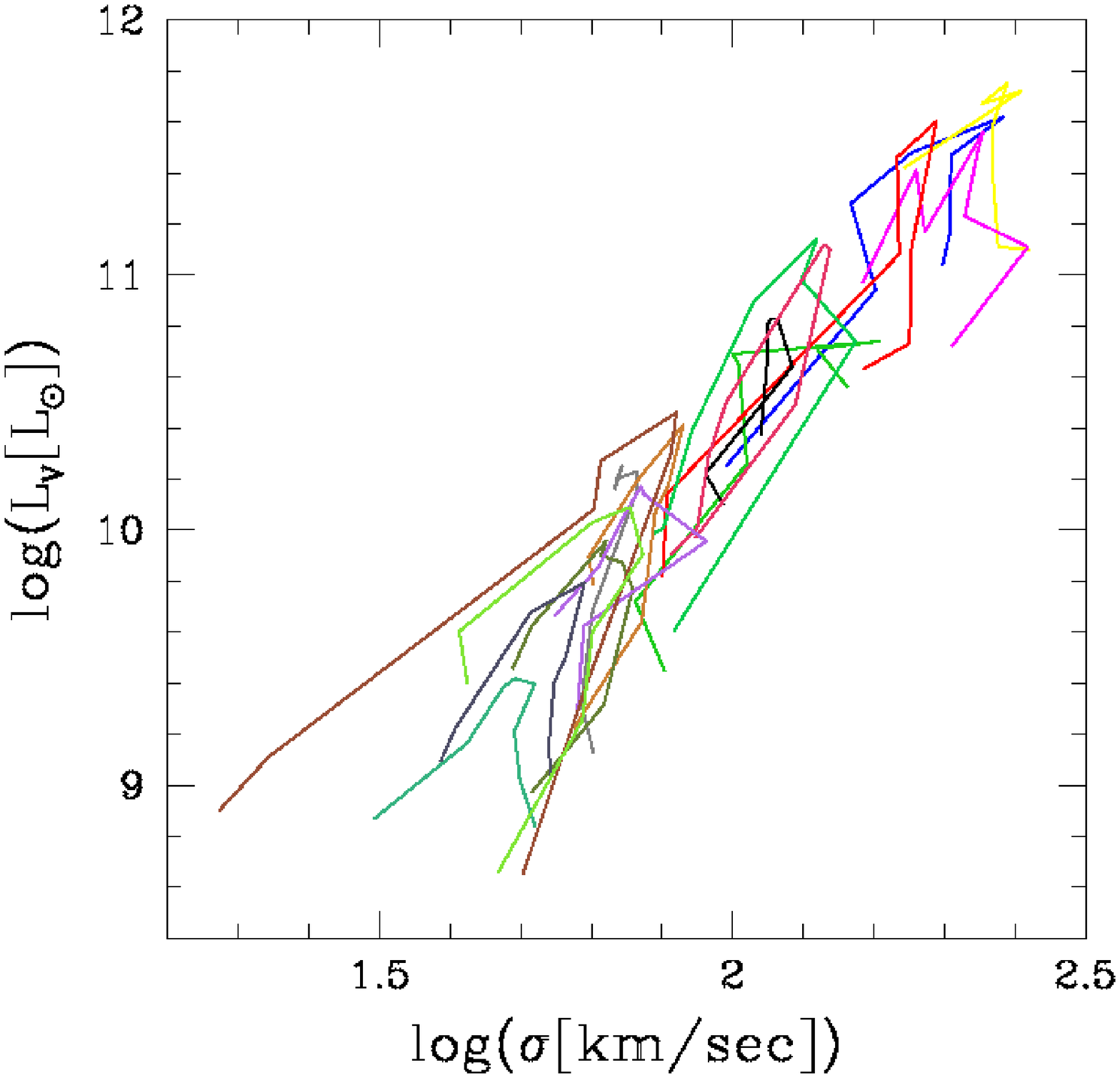}
   \includegraphics[scale=0.42]{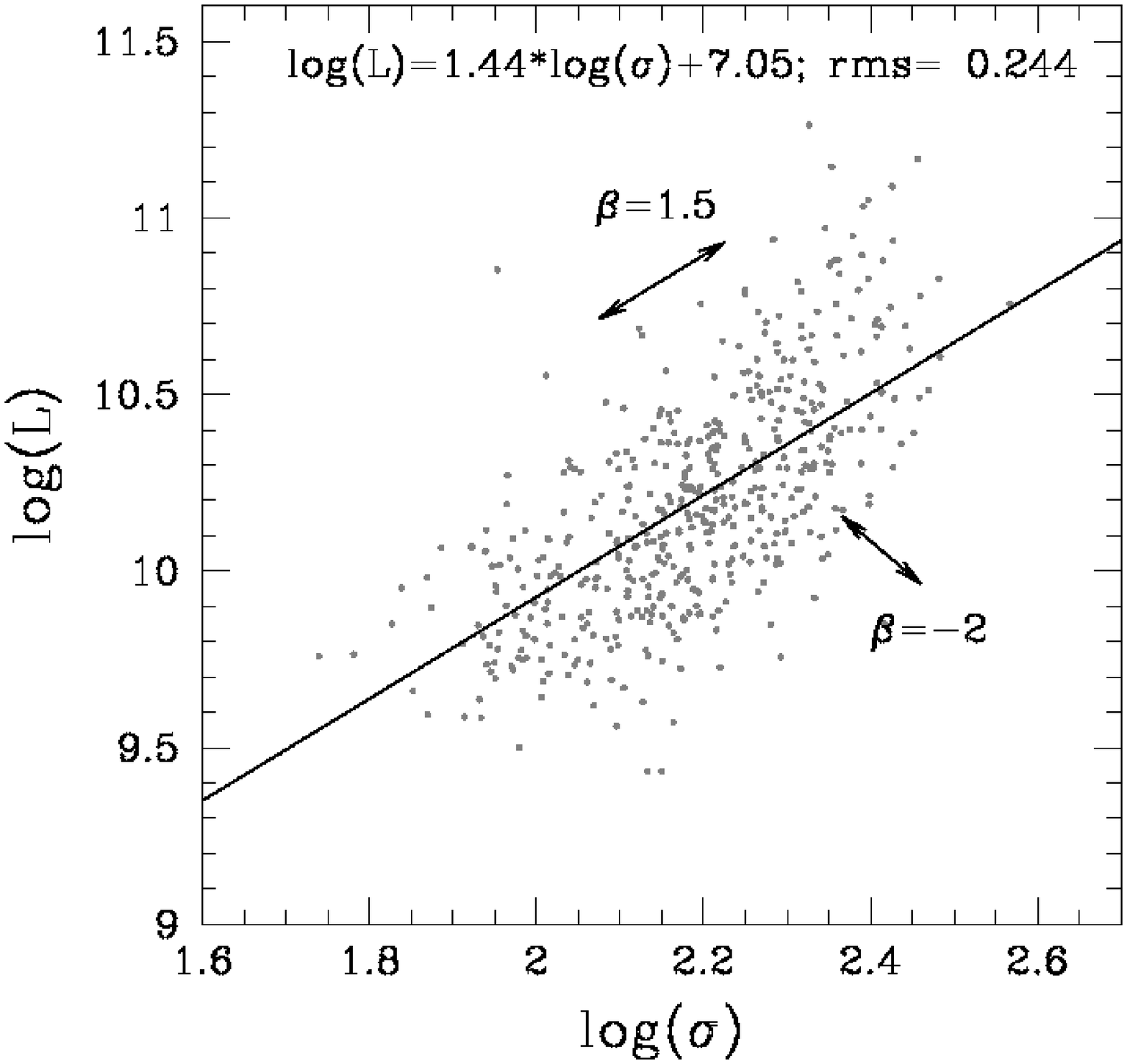}
   \caption{Left panel: the \Lsig\ plane filled by few simulated ETGs followed by the Illustris simulation from $z=4$ to $z=0$. Each color marks the path of a single galaxy. Right panel: the observed \Lsig\ plane. Gray dots are the ETGs with available $\sigma$ in our sample. The solid line is the least square fit. The arrow show the direction of motion for two different values of $\beta$.}
              \label{fig:1}
    \end{figure*}

\section{The FP and its projections}\label{sec:3}

The starting point of our approach originates from the behavior of ETGs in the \Lsig\ plane shown by the data of the Illustris simulation.
Figure \ref{fig:1} (left panel) gives an idea of the change of position of galaxies in this diagram during the evolution from $z=4$ to $z=0$.
In the figure we plotted with different colors a limited number of galaxies (of all morphological types), followed in their evolutionary path by the simulation. It is apparent from the figure that in the hierarchical framework galaxies change either luminosity and velocity dispersion { in both directions}, \ie\ both can increase or decrease according to the process at work (merging, stripping, quenching, gas accretion, star formation, etc.). 

We can account for  such behavior adopting  the law introduced by \citet{Donofrioetal2017} to explain the tilt of the FP:

\begin{equation}
    L = L'_{0}(t) \sigma^{\beta(t)}.
    \label{eq1}
\end{equation}

In this relation the parameter $L'_0$ and $\beta$ (both functions of time) can vary considerably from galaxy to galaxy, assuming both positive and negative values (see Fig. \ref{fig:1} right panel). Each galaxy might move in this plane across time, but only in the direction fixed by $\beta$ (marked by black arrows only for two possible cases).

This relation has nothing in common with the FJ law. It is a comfortable way for introducing the time variable parameters $L'_0$ and $\beta$, that the { Illustris simulations suggest}.
The advantage of this approach is that we are writing a relation that is now completely {independent of mass and  VT}, but includes time as hidden variable.

We will see below that the predicted change of position of a galaxy in this plane, as a consequence of evolution, has an important impact on the relations connected with the FP.

\begin{figure*}
   \centering
   \includegraphics[scale=0.8]{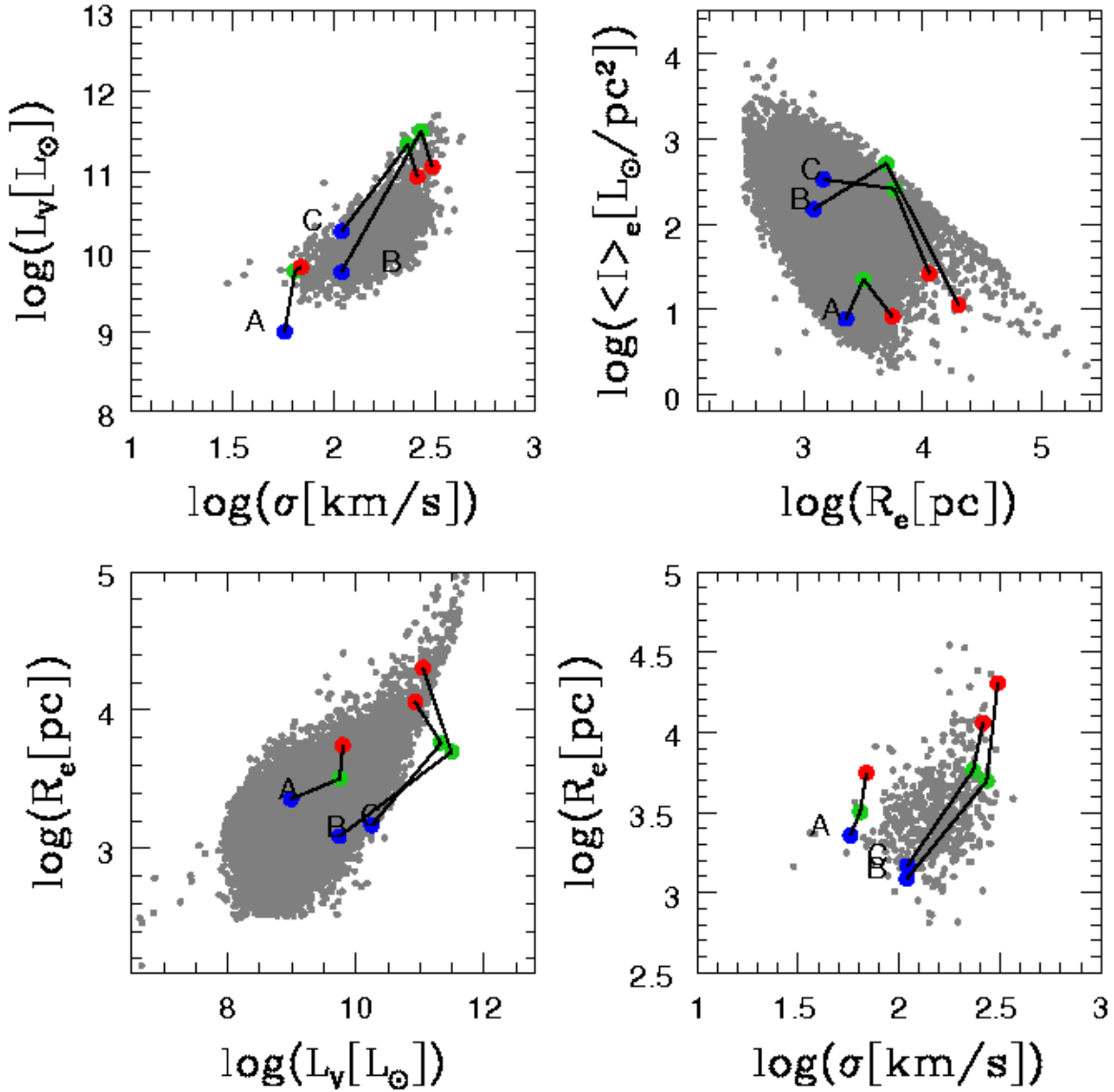}
   \caption{Four projections of the FP. The small gray dots mark the observational data. The colored dots are three simulated galaxies (A, b, C) of the Illustris simulation taken along their evolution respectively at $z=4$ (blue dots), $z=1$ (green dots) and $z=0$ (red dots). The solid lines connect each object in his evolution with time.}
              \label{fig:2}
    \end{figure*}

The Figure \ref{fig:2} shows the change of position for three galaxies of the Illustris simulation labeled A, B and C, starting from $z=4$ (blue dots) to $z=1$ (green dots) and $z=0$ (red dots), superposed to the real data (gray dots) at $z=0$ for all distributions linked to the FP. Each object is followed by the simulation in his transformation across time.

Note that while the \Lsig\ relation is almost preserved in shape and scatter (only a systematic tilt toward slower slopes is observed going from $z=4$ to $z=0$, \citet[see][]{Donofrioetal2020}), there is a substantial bending in all the other projections of the FP.

In general, the simulation follows quite well the trends observed in the real data (see below), reproducing in particular the clouds of points formed by the faint galaxies and the tails characterizing the distribution of the brightest galaxies.

The second equation of interest is that of the VT.
In the classical explanation of the FP tilt, it is implicitly assumed that a smooth variation of the zero-point of the VT equation, introduces the observed tilt. 
The VT in fact, can be written using luminosity instead of mass:

\begin{equation}
    L = 2\pi I_e R_e^2 = \frac{k_v}{G}\frac{L}{M_*}R_e\sigma^2
    \label{eq2}
\end{equation}
\noindent
where $G$ is the gravitational constant and $k_v$ is a term that gives the degree of structural and dynamical non-homology\footnote{This { allows us} to write $M_*$ instead of $M_T$.} as a function of the S\'ersic index $n$
($kv=((73.32/(10.465+(n-0.94)^2))+0.954))$ \citep[see][]{Bertinetal1992, Donofrioetal2008}).

The combination of eqs. \ref{eq1} and \ref{eq2} { leads}  to predict the slopes and zero-points in each projection of the FP as a function of $L'_0$ and $\beta$. We have indeed for the \IeRe\ plane:

\begin{equation}
I_e  = \Pi R_e^{\gamma}
\label{eq3}
\end{equation}
\noindent
where 

$$\gamma=\frac{(2/\beta)-(1/2)}{(1/2)-(1/\beta)}$$

\noindent
and
$\Pi$ is a factor that depends on $k_v$, $M/L$, $\beta$, and $L'_0$:

$$
\Pi  = \left [ \left (\frac{2\pi}{L'_0}\right )^{1/\beta} \left (\frac{L}{M_*} \right )^{(1/2)} \left (\frac{k_v}{2\pi G} \right )^{(1/2)} \right ]^{\frac{1}{1/2-1/\beta}}
\label{eq4}
$$

\noindent
This equation gives the only possible direction of motion of a galaxy in \IeRe\ plane as a function of $\beta$. The { orientation} of the motion depends on the type of transformation that a galaxy is experiencing (\eg\ merging, stripping, shock, star formation, quenching, etc.).

Another combination gives:
\begin{equation}
R_e = \left(\frac{1}{\frac{k_v}{G}\left(\frac{2\pi\langle I_e \rangle}{L'_0}\right)^{2/\beta}}\right)^{1/\left(4/\beta+1\right)}
M_*^{1/\left(4/\beta+1\right)}.
\label{eq5}
\end{equation}
\noindent
that is the \MRa\ relation. Note again { that both the slope and zero-point of the relation  for each single galaxy depend on $L'_0$ and $\beta$}.

In addition we have:
\begin{equation}
    R_e = \frac{k_v}{\Pi G}\frac{L'_0}{M_*}\sigma^{\frac{2+\beta}{1+\gamma}}
    \label{eq6}
\end{equation}
for the $R_e-\sigma$ relation, and

\begin{equation}
    R_e = \frac{G}{ k_{v} } \frac { M_{*} } {L} (L'_{0}) ^{2/\beta} L^{1-2/\beta}  
    \label{eq7}
\end{equation}
for the $R_e-L$ relation.

Finally we have:

\begin{equation}
    I_e = A \sigma ^{\frac{-\beta^3+6\beta^2-4\beta-16}{2\beta-4}}
    \label{eq8}
\end{equation}
\noindent
where 

$$
    A = \Pi^{(1-t)} \left (\frac{k_v}{G} \right )^\gamma \left 
    (\frac{L}{M_*} \right )^\gamma. 
    \label{eq9_a}
$$
\noindent
for the $I_e - \sigma$ relation.

Table \ref{beta_values} lists the values of the slopes { in the various projections of the FP predicted by   the values of $\beta$}.
Note that the values of these slopes implies a different motion for each galaxy, that is associated to the variation observed in the \Lsig\ plane.

Looking at this table in detail we can {note} that:
\begin{itemize}
    \item In all FP projections, when $\beta$ becomes progressively negative, \ie\ when the objects are rapidly declining in their luminosity at nearly constant $\sigma$, the slopes either converge to the values predicted by the VT (in the \IeRe\ relation and in the $R_e-L$ relation), or diverge toward large values (in the $I_e-\sigma$ and $R_e-\sigma$ relations), because the galaxy keeps its velocity dispersion when the luminosity decreases (only \Ie\ and \re vary);
    \item Both positive and negative values of the slopes (\ie\ the direction of evolution) are permitted, in agreement with simulations. Objects still active in star formation or experiencing a merger have positive values of $\beta$, while those {progressively quenching their stellar activity} have increasing negative $\beta$;
    \item The convergence of the slopes for progressively negative values of $\beta$, naturally { explains} the existence of the ZoE in the \IeRe\ and $R_e-L$ planes. Indeed, as $\beta$ decreases the galaxies cannot move in other directions.
    \item The curvature of the relations is naturally explained by the transition from positive to negative values of $\beta$.
\end{itemize}

\begin{table}
\begin{center}
\caption{The slopes in the log form of the \IeRe, \IeSig,  $R_e-M_*$, $R_e-L$, and $R_e-\sigma$ planes for different values of $\beta$.}
                \label{beta_values}
                \begin{tabular}{|r| r| r| r| r|}
\hline
         $\beta$     & \IeRe\ & \IeSig\ & $R_e$-$L$ & $R_e-\sigma$ \\
\hline
         4.0 &        0.0   &     0.0   &  0.50 &  6.00 \\  
         3.0 &        1.0   &    -0.50  &  0.33 &  2.50 \\
         2.0 &        -     &        -  &  -    &  4.00 \\
         1.0 &      -3.00   &     7.50  & -1.00 & -1.50 \\
         0.5 &      -2.33   &     5.54  & -3.00 & -1.87 \\
         0.0 &      -1.00   &     4.00  &  -    &  0.0  \\
        -0.5 &      -1.80   &     2.47  &  5.00 & -1.87 \\
        -1.0 &      -1.66   &     0.83  &  3.00 & -1.50 \\
        -1.5 &      -1.57   &    -0.98  &  2.33 & -0.87 \\
        -2.0 &      -1.50   &    -3.00  &  2.00 &  0.00 \\
        -2.5 &      -1.44   &    -5.24  &  1.79 &  1.12 \\
        -3.0 &      -1.40   &    -7.70  &  1.66 &  2.50 \\
        -3.5 &      -1.36   &   -10.40  &  1.57 &  4.12 \\
        -4.0 &      -1.33   &   -13.33  &  1.50 &  6.00 \\
        -4.5 &      -1.30   &   -16.51  &  1.44 &  8.12 \\
        -5.0 &      -1.28   &   -19.92  &  1.40 & 10.50 \\
        -8.0 &      -1.20   &   -45.60  &  1.25 & 30.00 \\
       -11.0 &      -1.15   &   -80.19  &  1.18 & 58.50 \\
       -25.0 &      -1.07   &  -360.35  &  1.08 &310.50 \\
       -50.0 &      -1.03   & -1347.92  &  1.04 &  1248 \\
      -100.0 &      -1.01   & -5197.96  &  1.02 &  4998 \\
     -1000.0 &      -1.00   &  -501998  &  1.00 &499989 \\
    -10000.0 &      -1.00   &-50020000  &  1.00 &50011512 \\
\hline
\end{tabular}
\end{center}
\end{table}

\begin{figure}
   \centering
   \includegraphics[scale=0.42]{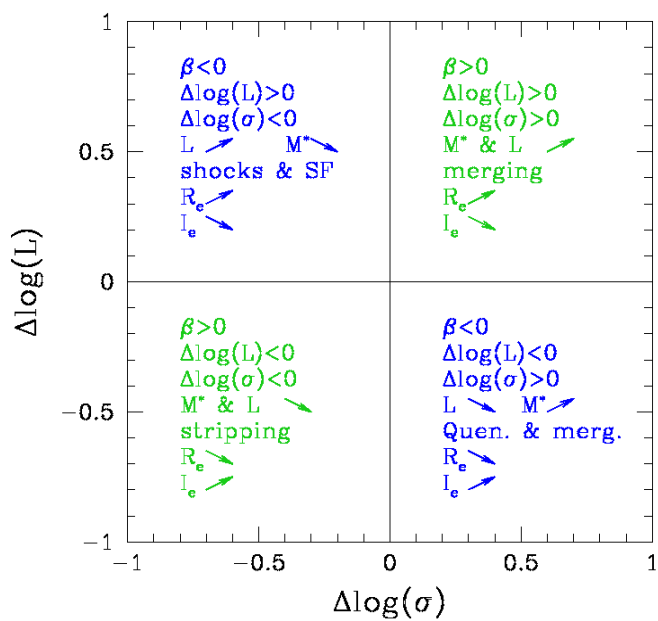}
   \caption{The $\Delta\log(L)-\Delta\log(\sigma)$ plane. The green and blue color mark the $\beta>0$ and $\beta<0$ directions of motion in the \Lsig\ plane respectively. Depending on this, a galaxy experiences a change in its structural parameters and move in the FP projections according to the slopes listed in Table \ref{beta_values} as described in the quadrant.}
              \label{fig:2b}
    \end{figure}

{
The easiest way to understand the  effects of all possible variations of the fundamental quantities defining the status of a galaxy  is to look  at the possible movements of a galaxy in the \Lsig\ plane during its evolution (four in total). Such movements are schematically shown in   
Fig. \ref{fig:2b}, which displays  according to the values of $\beta$,  the expected variations of \re, \Ie, $L$ and $M_*$, as indicated by the arrows. The figure also indicates the possible physical mechanism at the origin of the observed variation. Panels with  green and blue colors correspond to positive and negative values of $\beta$. Note that when $\beta$ is negative, not necessarily there is a decrease in luminosity while when $\beta$ is positive, a decrease in luminosity may occur.} 
{ When the luminosity changes, the effective radius (\ie\ the radius that encloses half the total luminosity) changes, as well as the mean effective surface brightness \Ie.}

\begin{figure*}
   \centering
   \includegraphics[scale=0.42]{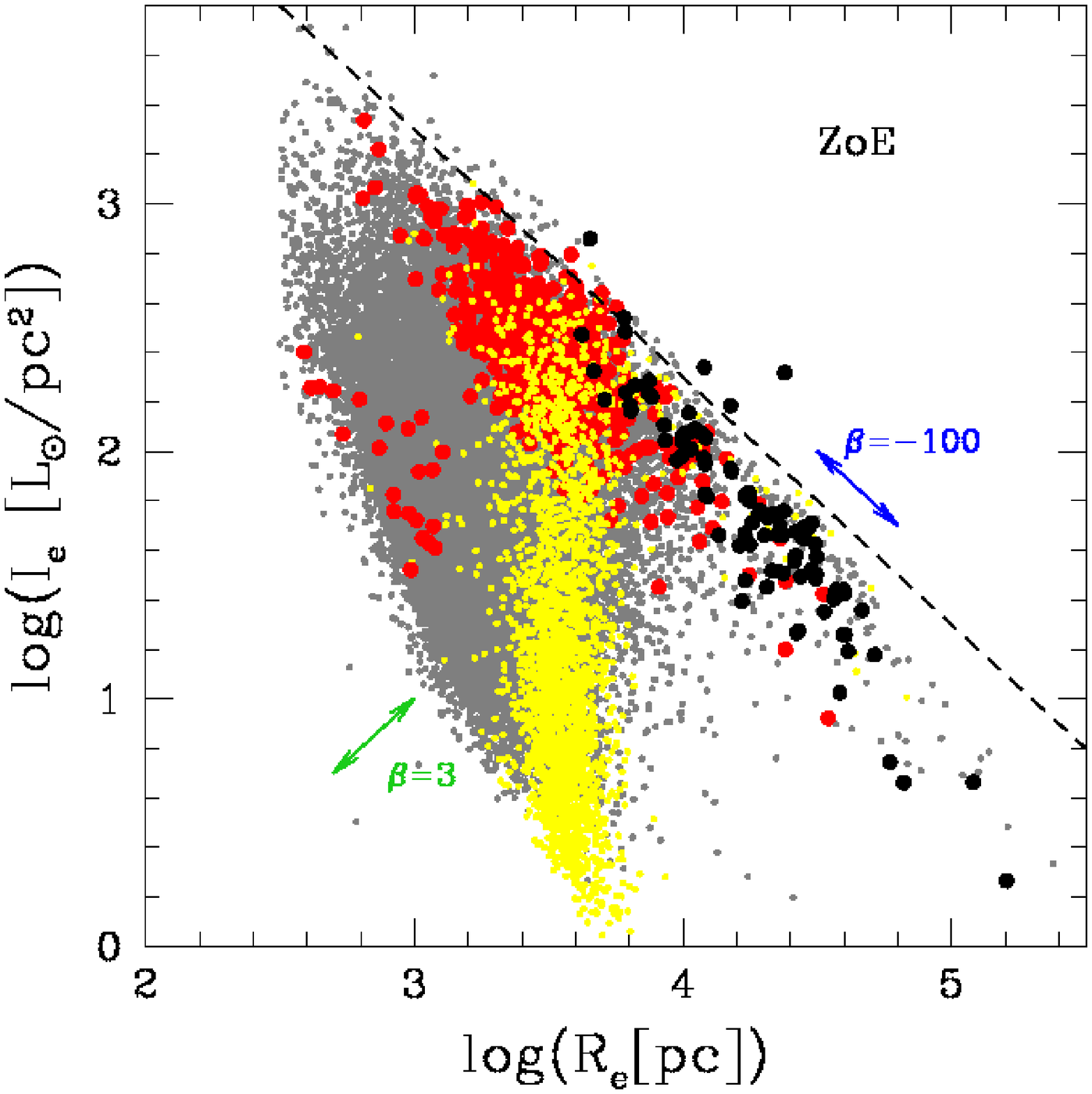}
   \includegraphics[scale=0.42]{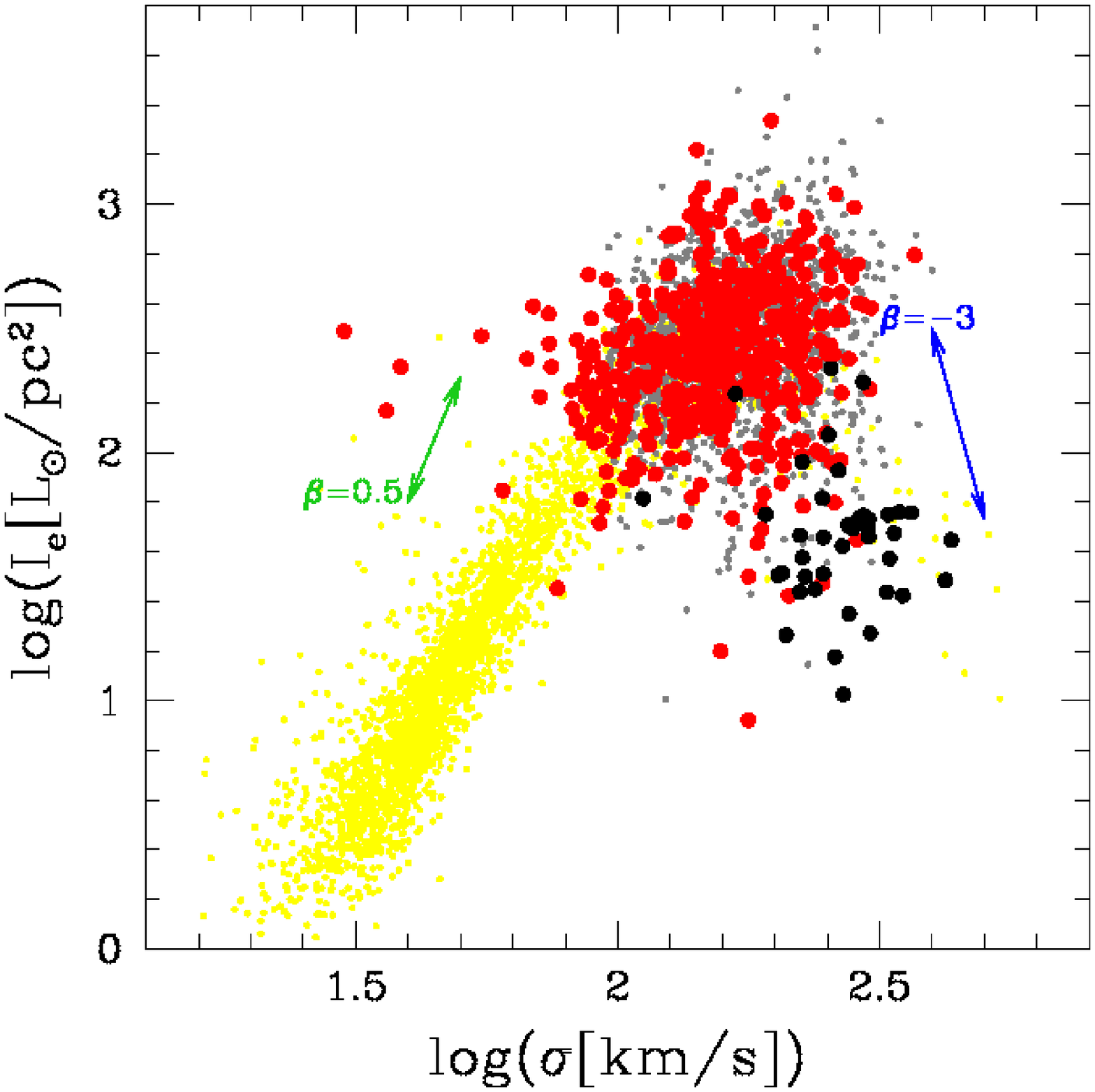}
   \includegraphics[scale=0.42]{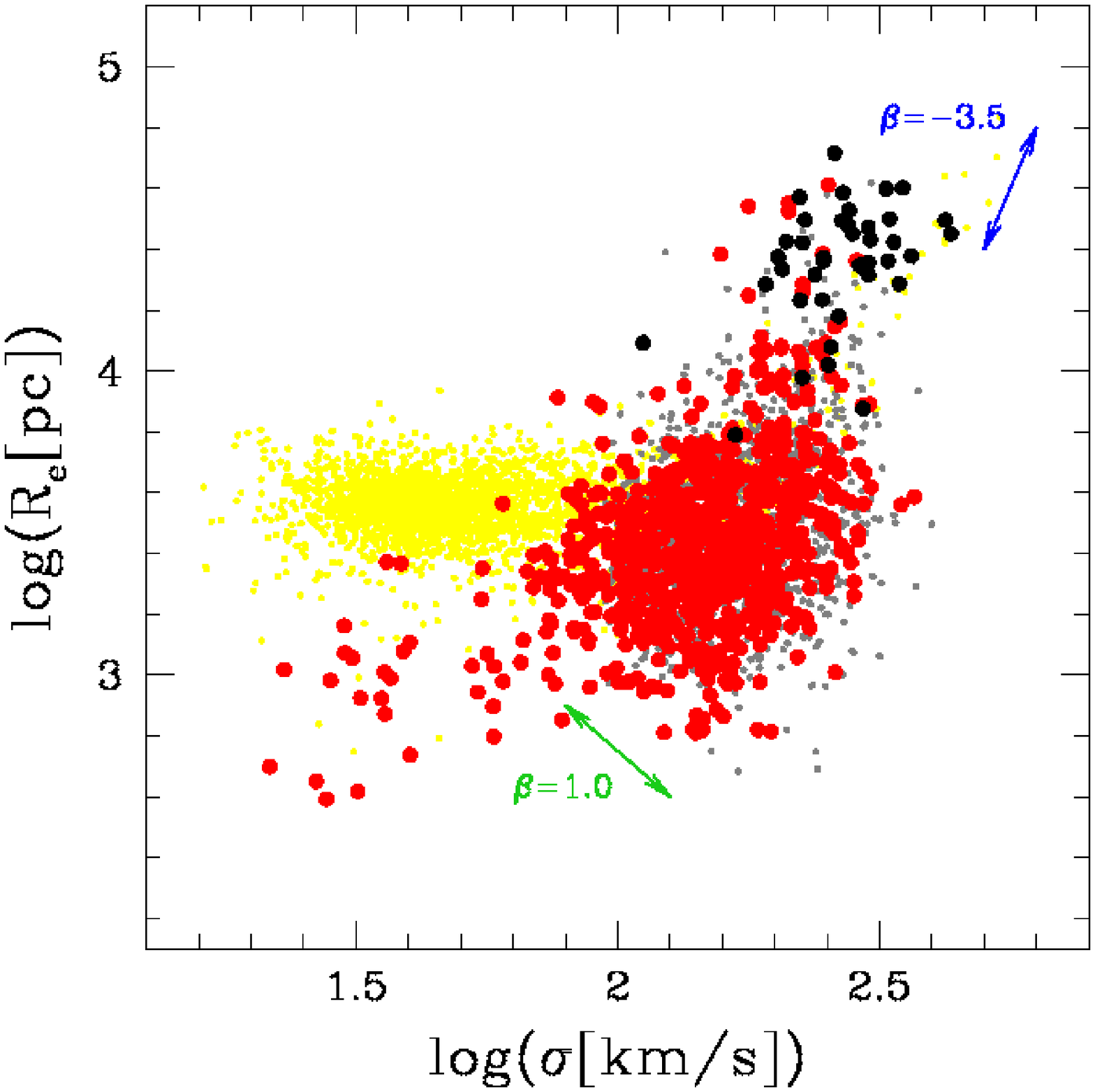}
   \includegraphics[scale=0.42]{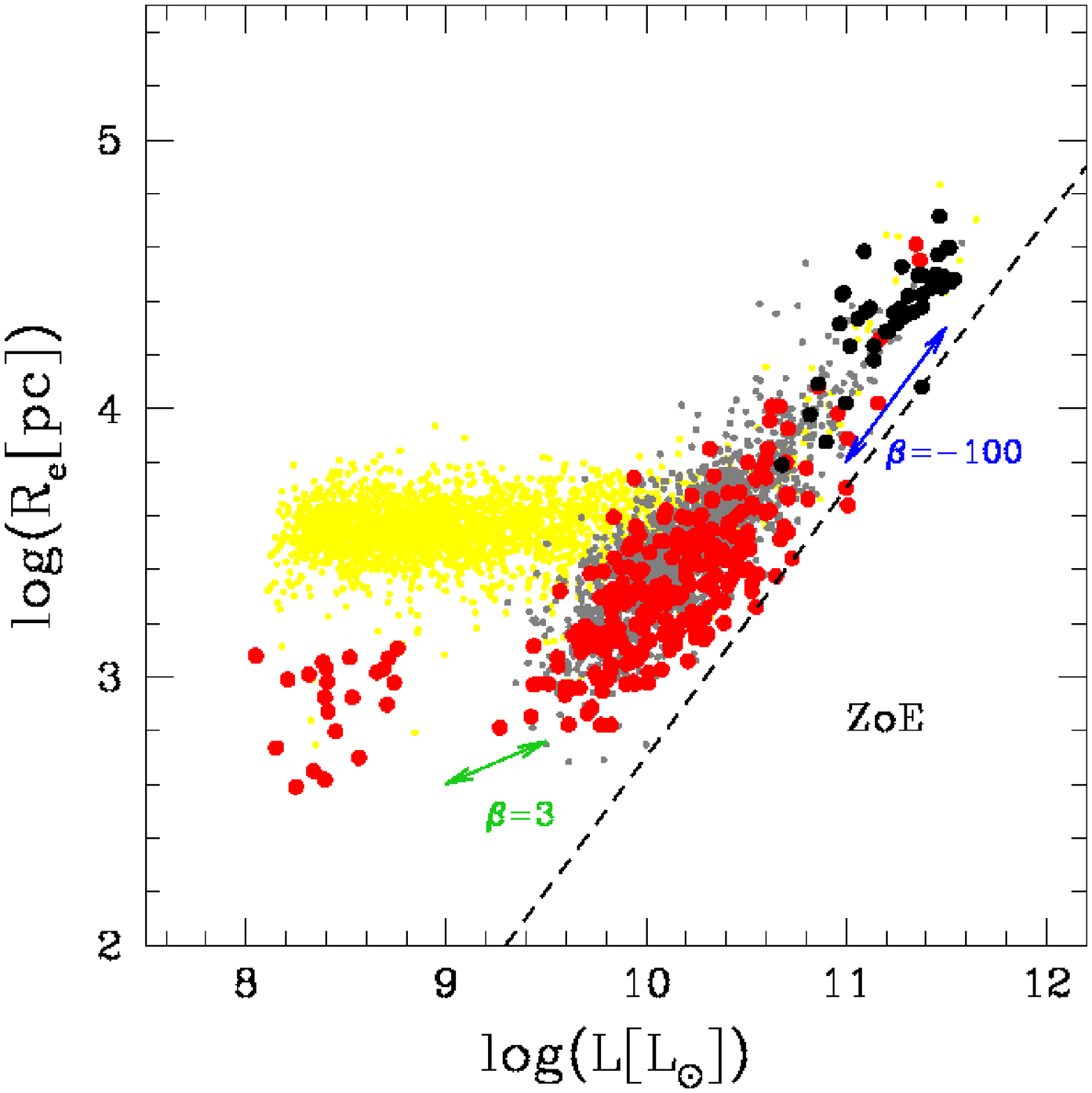}
   \caption{Four different projections of the FP. The yellow dots mark the data of the Illustris simulation. The red dots are the sample with available stellar masses and velocity dispersion. The black dots are the BCG and 2nd brightest ETGs of the clusters. The gray small dots are the optical and spectroscopic data for the objects that have not a measured stellar mass. The dashed lines mark the region of the ZoE. The green and blue arrows mark the direction of a future evolution expected on the basis of the values of $\beta$. The green (blue) color indicates a positive (negative) value of $\beta$.}
              \label{fig:3}
    \end{figure*}

{All these effects are better illustrated}  in Fig. \ref{fig:3}, which shows the projections of the FP at redshift $z=0$ (the present). In each panel we also visualize the direction that each galaxy would follow due to the temporal evolution of the fundamental relationship  \Lsigbtempo\  via the parameters $L'_0$ and $\beta$, this latter in particular. The virtual displacement for negative (positive) values of $\beta$  is indicated by the blue (green) arrows. Common to all panels is the striking  curvature of all the {relations} and the existence of ZoE in two of them that seems to correspond to large negative values of $\beta$. 

{ Before examining this figure in great detail, it is worth recalling once more that any change of the luminosity (increase or decrease), while the luminosity profile across the galaxy is kept fixed, will immediately  mirror on the radius $R_e$ and surface brightness $I_e$.
For instance, a decrease in luminosity by a factor of two will decrease $R_e$, but increase $I_e$. The opposite for an increase of $L$. In general this correlation among the three variables  is preserved for any arbitrary variation of the luminosity. The second thing to keep in mind is that $R_e$ does not strictly correspond to the radius of the VT, it is proportional to but not identical to it; consequently in a real galaxy, $L$, $I_e$ and $R_e$ may change with time, but $\sigma$ may remain constant.  Last thing to remember is that the slope of the arrows displayed in each panel visualizes the expected slope of the displacement of a generic galaxy based on the value of $\beta$ labeling each arrow according to the entries in Table  \ref{beta_values}. 
Finally, it should be emphasized that the arrows  indicate the direction not the orientation  of the future temporal evolution of a galaxy. Furthermore, they are not the path  followed by  each galaxy to reach  the current observed position.
}

{
(i) In the \IeRe\ plane, moving across from left to right,  $\beta$ gradually changes from positive to negative values, and the slope of the associated displacement vectors gradually changes from $1$  ($\beta=3$) to $-1$  ($\beta = -100$), \ie\ toward the value predicted by the VT. The cloud of points with $\log(R_e)<4$ is likely due to a mixture of possible transformations of galaxies (with positive and negative values of $\beta$) determining different effects  on \Ie\ and \re, as a consequence of changes in luminosity. For $\log(R_e) > 4$, the tail of the bright galaxies develops, which is populated  by  objects that nowadays have negative $\beta$ and consequently can move only along the direction with slope to $-1$. It is important to emphasize that the observed \IeRe\ plane can only catch the situation at $z=0$ as photographed by the  data. Nevertheless, thanks to the effect of $\beta$ on the slope of the various relations (see Table  \ref{beta_values}), we have an idea of the effect of time  on the position of galaxies in this plane via the effects on $L$, $I_e$, and $R_e$.  In this plane the ZoE is also  visible, the border of which is the  dashed line. The region at the right hand side of this line is forbidden to  galaxies. In the current view, this region arises naturally when the values of $\beta$ become large and negative ($\beta=-100$) which  yield a slope for the boundary equal to $-1$, i.e.  the same value one derives from the VT. Galaxies that are fully dynamically relaxed and in a phase of passive evolution at decreasing luminosity  must fall close to the {ZoE} boundary from the side of the permitted area.

(ii) The $I_e-\sigma$  is a second projection of the FP. Note the strong bending of the ETG distribution in such diagram. Again, the values of $\beta$ predict slopes of the corresponding displacement vectors that fairly account for the distributions of the observational data in  this plane. Clearly, the surface brightness of an object that is decreasing its luminosity, rapidly falls down at nearly constant $\sigma$ (see the position of the BCGs). On the other hand, when the evolution predicts an increase in luminosity, the galaxies climb the relation. Remember that a negative value of $\beta$ does not necessarily mean a decrease in luminosity. The simulations of model galaxies at different redshifts (times) show that an increase in $L$ is possible even when $\sigma$ decreases. Such an event may likely happen \ie\ as a consequence of stripping, when some stars are removed from the galaxies, but a shock induces a new star formation event that increases the galaxy luminosity. We can also appreciate that negative values of $\beta$ reproduce quite well the tail of  BCGs. Note also that in this diagram there is no clear evidence of the ZoE, \ie\ the values of $\beta$ do not converge to a limit, as in the case of the \IeRe\ relation. When $\beta$ starts to be negative, there is a progressive decrease in the mean surface brightness due to the smaller luminosity. Consequently, the effective radius,  can only decreases at nearly constant $\sigma$ along the direction permitted by $\beta$. Finally, we note that the Illustris simulations for low mass galaxies (yellow dots) largely miss the the position of real objects.
Since in the \IeRe\ plane the artificial galaxies have systematically 
larger \re, with respect to real galaxies, here their surface brightness is systematically lower. However the tail of the BCGs is again well reproduced.
The explanation could be the same suggested by \citet{Chiosietal2020} for the $R_e$-$M_*$ relation, specifically bad estimates of the energy feed-back and presence of galactic winds. 

(iii) {The $R_e-\sigma$ plane is the third projection of the FP. The diagram shares many points in common with the previous one, specifically  the central bulk of ETGs (the crowed cloud), the tail formed by the BCGs with large $R_e$ and $\sigma$ characterized by $\beta=-3.5$ and the region of the faint dwarfs. The values of $\beta$ in the cloud and in the region of the dwarfs might be either positive or negative. In the figure we plotted the value of $\beta=1$ that indicates a movement nearly orthogonal to that followed by the bright galaxies. The same remarks made above about the missing evidence of ZoE and discrepancy between theory and data concerning  the position of the low mass galaxies can be applied.}

(iv) The $R_e-L$ plane. With the variables $I_e$ and $R_e$ we can construct another plane (the $R_e$-$L$ plane) which is a proxy of the well known $R_e$-$M_*$ plane, because the luminosity of a galaxy is mainly proportional to the stellar mass (another possible source of luminosity variation, but much less important). But for the period of time with intense star forming activity, during which the luminosity is very high and changes on a short time scale, when the star formation is over, the luminosity gently and steadily decreases. Differences in chemical composition induce modest variations in the luminosity; merger episodes (unless among objects of comparable mass) causes variations in luminosity of moderate intensity that are soon smeared  by the dominant underlying stellar populations. 
In general, the Illustris simulation is able to reproduce the curvature of the relation:  faint galaxies distribute with a mean slope (about 0.3) that correspond to $\beta\sim3$,  ETGs roughly distribute with mean slope 0.5 ($\beta=4$), and finally BGCs are located in a tail with slope $\simeq 1$ and $\beta<0$ (approximating the same value predicted by the VT). These objects are indeed very old and { do } not suffer anymore the effects of merging and star formation. They are evolving passively. 
In this plane we can identify again the { ZoE} and its boundary along which the parameter $\beta$ is high and negative and the slope of the $R_e$-$L$ relations tends to 1.

In summary, what we claim here, is that all these diagrams should be analyzed taking into account the effects of time. They are snapshots of an evolving situation, and such evolution cannot be discarded in our analysis. The \Lsigb\ law { catches  such evolution in a correct way}, predicting the correct direction of motion of each galaxies in its future. This way of reasoning permits us to understand why the galaxies are in the positions we observe today. {In other words the scaling relations become tools to infer the evolutionary status of each galaxy}.

A further possibility offered { by the joined action of the VT and \Lsigb\ relation together} is to combine eq. \ref{eq3} and \ref{eq6} obtaining an expression for the FP as a function of $\beta$. We get:

\begin{equation}
    a_1 \log\sigma + b_1\log I_e + c_1 \log R_e + d_1 = 0
    \label{eq9}
\end{equation}

\noindent
where

$$a_1 = (\beta^3-4\beta^2+8)$$
$$b_1 =-2(\beta-2)$$
$$c_1=-4(\beta-3)$$
$$d_1=+2(\beta-2)\log(k_v/G) - 2(\beta-2)\log(M_*/L).$$

Note that in the hierarchical framework all the FP coefficients depend on $\beta$. In addition, it is remarkable that the coefficients $a$, $b$ and $d$ of eq. \ref{eq9} are exactly 0 when $\beta=2$. This is due to the fact that for this exponent, the VT and the \Lsigb\ law are the same relation. The combination of the two relations in this case has no meaning.

Conversely we can derive $\beta$ from eq.\ref{eq9}, being this parameter the only unknown.
{ Indeed inverting  eq. \ref{eq9} we can derive $\beta$ for each of the 479 galaxies of the WINGS sample.}
We get:
\begin{equation}\begin{split}
    \beta^3\log\sigma - 4\beta^2\log\sigma+ \\
    - 2\beta(\log I_e + 2\log R_e - \log(k_v/G) + \log(M_*/L)) + \\
    + 8\log\sigma + 4\log I_e + 12 \log R_e - 4 \log (k_v/G)+ \\
    4\log(M_*/L)=0
    \label{eq10}
    \end{split}
\end{equation}
\noindent
which is a cubic equation in $\beta$ of the type $$\alpha_1\beta^3+\alpha_2\beta^2+\alpha_3\beta+\alpha_4=0$$
with coefficients:
$$\alpha_1=\log\sigma$$ 
$$\alpha_2=-4\log\sigma$$ 
$$\alpha_3=-2(\log I_e + 2\log R_e - \log(k_v/G) + \log(M_*/L))$$  $$\alpha_4=8\log\sigma + 4\log I_e + 12 \log R_e - 4 \log (k_v/G)+4 \log(M_*/L).$$

This equation admits three real solutions when the discriminant $\Delta < 0$\footnote{This is the case for many of our solutions, but a small change in the input coefficients $\alpha$ might result in complex solutions (see text).}. The mean value of the solution n.1 for our sample of ETGs is $<\beta_1>=2.78\pm0.08$, that of solution n.2 is $<\beta_2>=-2.54\pm0.06$, and that of solution n.3 $<\beta_3>=3.76\pm0.07$. It is therefore interesting to note that despite the differences in $\sigma$, \re, \Ie, $n$ and $M_*/L$ among galaxies, the scatter around each solution is very small. As a matter of fact,  { the mean values of the coefficients $\alpha$ for the observational values of  the $I_e$, $R_e$, $M_*$, $k_v$, and $\sigma$ of the 479 ETGs, are}:

$$<\alpha_1>=2.18\pm0.15$$ 
$$<\alpha_2>=-8.74\pm0.59$$ 
$$<\alpha_3>=-14.00\pm0.89$$
$$<\alpha_4>=59.51\pm3.39.$$
whose uncertainties are not negligible. Indeed, considering the $3\sigma$ interval of possible variation of the $\alpha$ values, it is somewhat surprising to find three solutions for $\beta$ with such small scatter.}

We can only suspect that in this sample of ETGs, at variance with the large possible values of $\beta$ permitted by simulations (see below),
only well defined values of $\beta$ are permitted and notably these are both positive and negative in agreement with simulations.

The only possible explanation for such fine-tuned values of $\beta$ is that all the ETGs considered here have had very similar histories of mass assembly. The general impression is that there are on average two groups of galaxies, one with negative values of $\beta$ that are likely today in quenched state, and one with positive $\beta$ { and therefore} still active in their evolution. Notably, large negative values of $\beta$ are not observed, a fact that implies that the galaxies have not yet reached the pure passive evolution, when there are no more mergers and other disturbing effects. 

Unfortunately, we cannot say which is the correct value of $\beta$ for each galaxy. {Nevertheless  We will see below that some solutions can be discarded on the basis of the constraints imposed by the FP}.

\begin{figure}
   \centering
   \includegraphics[scale=0.42]{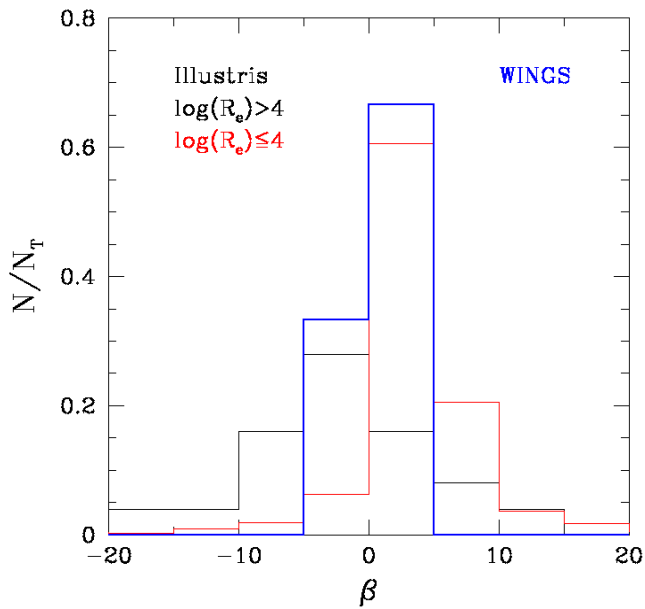}
   \caption{The histogram of the calculated values of $\beta$ is marked with a blue line. The values of $\beta$ from the Illustris simulation are shown in two histograms with black and red colors to distinguish objects with $\log R_e>4$ and $\log R_e \leq 4$ respectively.}
              \label{fig:4}
    \end{figure}

Figure \ref{fig:4} shows the solutions we have obtained for $\beta$ compared with the values derived from the Illustris simulation. In that case $\beta$ was easily obtained by looking at the values of $L$ and $\sigma$ at two different redshift epochs: $$\beta=\Delta\log(L)/\Delta\log\sigma.$$ 
{In the figure we estimated $\beta$ passing from from $z=1$ to $z=0$.}

Note the impressive superposition of these histograms, obtained in two completely different ways, and the fact that both negative and positive values are admitted as solutions. The simulation however, predicts the existence of both larger negative and larger positive values of $\beta$ for few galaxies \citep[see,][]{Donofrioetal2020}, a fact not observed in our data.
This is possibly due to the fact that in the simulations all types of objects are considered. The smaller ones in particular can be { characterized by  to large positive and negative variations of $\beta$ (star formation events, merging, big stripping of material, etc.). What it is important to stress, is that the large majority of the $\beta$s, predicted by the simulations, are in the same region predicted by our calculations.}

We provide here a possible estimated error for the $\beta$ solution. We get: $\Delta\beta_1=\pm0.16$, $\Delta\beta_2=\pm0.016$ and $\Delta\beta_3=\pm0.35$. These have been obtained by adding the known uncertainties on $\sigma$, \re, \Ie, $n$ and $M_*/L$ and deriving the new solutions for $\beta$. However, in some cases, when we add such uncertainty to the structural parameters, we obtain many complex solutions for $\beta$. On the contrary, when we drop these uncertainties we get again three real solutions for $\beta$. This fact requires a much deep analysis of the $\beta$ solutions and it is postponed to future studies. What is clear, is that the possible real solutions for $\beta$ requires a very good knowledge of the structural parameters of the galaxies. Indeed, the possibility of determining $\beta$ (and therefore $L'_0$) for each galaxy, is { rich} of possible consequences for our study of galaxy evolution. What values of $\beta$ could be obtained for high redshift galaxies?

{ An important fact to stress now instead, is that knowing  the solutions for $\beta$ immediately leads us to  determine the zero-point $L'_0$ on the basis of eq. \ref{eq1}.}

  \begin{figure}
   \centering
   \includegraphics[scale=0.42]{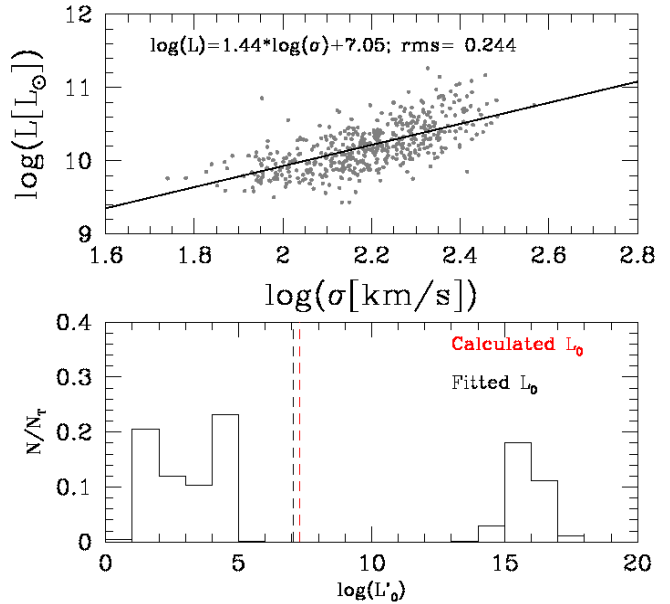}
   \caption{Upper panel: the FJ plane. The gray dots are the observational data and the solid black line the least square fit. Bottom panel: The histograms of the calculated values of $L'_0$ starting from the solution for $\beta$. The average value of $L'_0$ considering all the solutions is marked by the red dashed line, while the value obtained from the fit of the FJ relation is marked by the black line.}
              \label{fig:5}
    \end{figure}

The figure \ref{fig:5} clearly demonstrate that the zero-point $L_0$ of the FJ relation, that is given by the fit of the ETGs distribution in the \Lsig\ plane, is simply the { average value} of the large number of $L'_0$ obtained for each galaxy. Within the errors the average of $L'_0$ and the fitted value $L_0$ are indeed identical. It follows that the FJ relation is the average distribution resulting naturally from the hierarchical evolution of ETGs.

\begin{figure}
   \centering
   \includegraphics[scale=0.42]{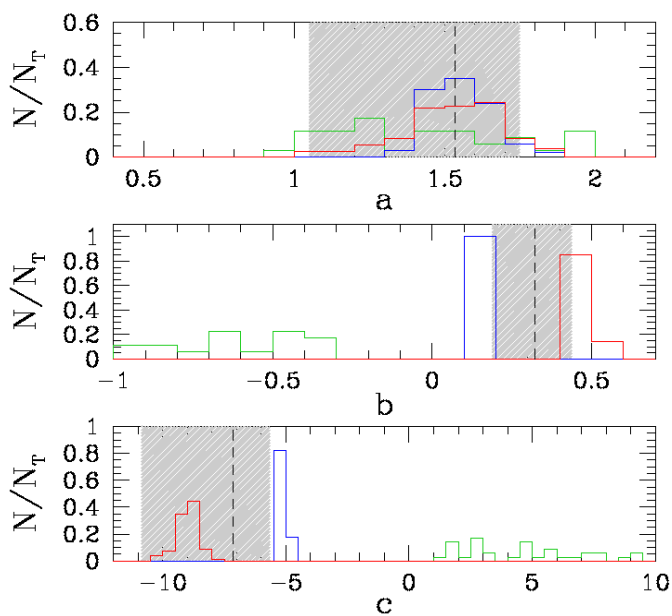}
   \caption{Histograms of the values of the FP coefficients $a$, $b$ and $c$. The green, blue and red colors correspond to the three solutions found for $\beta$ respectively. The gray band in each plot marks the region of the coefficients obtained by \cite{Donofrioetal2008} for the fit of the FP of the different WINGS clusters. The dashed black lines are the average values of the coefficients obtained considering only the solutions $\beta_2$ (blue) and $\beta_3$ (red), discarding the solution $\beta_1$ (green) that clearly does not fit with the intervals of the measured FP coefficients.}
              \label{fig:6}
    \end{figure}

The final step, shown in Fig. \ref{fig:6}, is that related to the
FP coefficients. Eq. \ref{eq9} tells us that the FP coefficients are a function of $\beta$. In the figure the gray bands mark the interval of $a$, $b$ and $c$ obtained by \citet{Donofrioetal2008} fitting the FP separately for each cluster of the WINGS data-set. The histograms mark the distribution of the coefficients $a$, $b$ and $c$ obtained writing the FP in its classical form:
$$\log(R_e)=a\log(\sigma)+b<\mu>_e+c,$$
where the units are in kpc for \re and $mag/arcsec^2$ for $<\mu>_e$.

The plotted solutions are in different color according to the values
of $\beta$. Notably, we observe that the solution $\beta_1$ does not fit the $b$ and $c$ coefficients. For this reason we decided to discard such solution for the moment.

Again we see that the mean value of the coefficients obtained from the $\beta_2$ and $\beta_3$ solutions, that are respectively all negative and positive, perfectly enters in the gray region of the permitted coefficients measured by the fit of the FP.

The correct expression of the FP coefficients as a function of $\beta$ is further { secured} by the fact that, when $\beta=4$, \ie\ when we use the value of the FJ relation predicted by the VT, the values of the calculated coefficients of the FP as a function of $\beta$, written in the form $\log R_e = a \log\sigma + b\langle I\rangle_e +c$ (with \re\ in kpc and \muem\ in $mag \,arcsec^{-2}$) are exactly the values predicted by the VT, \ie\ $a=2$ and $b=0.4$.

In summary we claim that, also in this case, the values of the coefficients derived from the fit of the FP, are nothing else that the average value of the two possible solutions for the coefficients found for each $\beta$. Clearly, each galaxy sample under analysis will have its own number of objects with $\beta_2$ and $\beta_3$ values (in agreement with the evolutionary status of the sample). This naturally accounts for the small variations observed in the fits of the FP obtained from different data samples.\footnote{We remark however, that the large interval of the FP coefficients found by \cite{Donofrioetal2008} are also due to the small number of galaxies available for some clusters. Clusters with less than 30 objects have a large error in the measured FP slope.}

It is also important to remark that, as seen in the middle and bottom panel of Fig. \ref{fig:6}, there are no values for $\beta$ that can reproduce the most accredited values for the coefficients $b$ and $c$, ($\approx0.3$ and $\approx-8$ respectively). This means that the measured values coming from the fit of the FP are averages of the possible coefficients for $b$ and $c$. These are different because there are galaxies with negative and positive values of $\beta$.

In addition, the existence of different coefficients as a function of $\beta$ might naturally account of the claimed bending of the FP, when we insert in the sample the small ETGs. In other words, the FP is not a plane, but a surface, originating from the hierarchical evolution.

Having demonstrated that the combination of the VT and \Lsigb\ law is the key to interpret the different aspects of the FP projections, we can discuss now how it possible to account for the variation of the slope visible in the FJ relation and measured by several authors for the bright ETGs.
Indeed, starting from the VT:

\begin{equation}
    \sigma^2 = \frac{G}{k_v}\frac{M_*}{R_e},
\end{equation}
when the luminosity is substituted to $M_*$ and $R_e$, one obtains two different trends:

\begin{center}
\begin{eqnarray}
 L = \frac{k_v}{G}\, \sigma^{2.10}     & {\rm when}\, M_* \leq 10^{10} M_\odot \\
\smallskip
 L = \frac{k_v}{G}\, \sigma^{4.76}     & {\rm when}\, M_*   >   10^{10} M_\odot 
\end{eqnarray}
\end{center}

These values of the slopes, different for high and low luminous galaxies, are in good agreement with observations \citep[see \eg][]{Choietal2007,Desrochesetal2007,HydeBernardi2009,Nigoche-Netroetal2010,Montero-Dortaetal2016}.

This change of slope occurs because $M_*$ { scales} as $L^{1.12}$ over the whole mass range, while $R_e$ follows two different trends at the low and high mass ranges. In particular we derive that $R_e\propto L^{0.17}$ at the low masses and $R_e\propto L^{0.70}$ at the high masses. The slopes in log units are therefore very close to that observed. Clearly a small difference in the $R_e - L$ relation can fit exactly the trends derived from the fits. 

The last thing to discuss is the origin of the small scatter visible both in the FJ and FP relations. In the Illustris numerical simulation the intrinsic scatter around the FJ relation is always quite small ($\approx0.09$), despite there is a clear rotation of the whole distribution towards smaller slopes ($\approx 2$) going from $z=4$ to $z=0$ \citep[see,][]{Donofrioetal2020}. 
The same effect has been observed for the FP \citep{Luetal2020} with the {Illustris-TNG} data. As discussed {in the introduction} the most accepted view is that galaxies deviate from the plane according to their $M/L$ ratio.

In the hierarchical context the concept of FP needs a clarification. We need to remark that the FP is only the fit of a distribution. In reality each galaxy follows its own FP relation, given by eq. \ref{eq9}. The measured coefficients are only averages of the possible FP coefficients of each galaxy. Consequently, the scatter is also a not well defined concept in this case. Since there is not a unique relation, that {allow us}  to define the scatter around it, we can only try to reproduce the scatter in this context by supposing that all galaxies have the same $\beta$. Indeed, in this case the coefficients $a_1$, $b_1$, $c_1$ of eq. \ref{eq9} are all equal, while $d_1$ can change for the different values of $k_v$ and $M_*/L$. In particular if we put $\beta=0.6$ (the mean value of $\beta_2$ and $\beta_3$) the observed scatter is $\approx0.06$\footnote{This solution for $\beta$ does not reproduce however the observed FP.}, a value very close to the measured intrinsic scatter of the FP. This means that the deviations from the FP have been correctly attributed to the non-homology of galaxies and their different mass-to-light ratio.

The small observed scatter implies only that
the ETGs are a very homogeneous population of galaxies, with very similar properties on average. The hierarchical evolution does not alter the scatter of the FP, as in the case of the FJ relation.
In general we can argue that the main reason for such small scatters in the FJ and FP relations is necessarily linked to the role of gravity. Gravitation provides the dynamical relaxation of the stellar systems in short times with respect to the time scale of stellar evolution. The galaxies are always close to the virial condition. Luminosity, on the other hand, can vary in the way we have seen, but never with increments/decrements larger than a factor of $2-3$, \ie\ about $0.2-0.3$ in log units. It follows that the scatter around the FJ and FP relations remains always very small, even if the luminosity variations operate the changes in the aspect of the FP projections.

\section{Conclusions}\label{sec:4}

In summary, the use of the \Lsigb\ relation as a proxy of evolution, marking the path followed by ETGs in the \Lsig\ plane, when combined with the equation of the VT, provides the following evidences:

\begin{enumerate}
    \item The FP can be expressed with coefficients that are function of $\beta$. This means that the FP must evolve with redshift and that the coefficients must depend on { the adopted waveband in which observations are taken as confirmed by the current observational data};
    \item Only two of the three possible real solutions found for $\beta$ (using the observations of the WINGS database) provide the correct coefficients of the FP. {Furthermore, the bending of the FP might naturally be explained by invoking different solutions for $\beta$};
    \item The tilt of the FP results from the average values of the FP coefficients of eq. \ref{eq9} as a function of $\beta$;
    \item The zero-points of both the FP and FJ relations are the average values of the coefficients obtained for different $\beta$;
    \item All the characteristics of the FP projections can be explained at the same time. This includes: 1) the curvature of the relations, that turns out to depend on the existence of positive and negative values of $\beta$, and 2), the existence of the ZoE, \ie\ the line marking the separation between the permitted and not permitted regions in these planes. No galaxies can reside in the ZoE. 
    \item {The ZoE is obtained in a natural way when the evolution start to quench star formation in the galaxies in such a way that their luminosity progressively decreases. This corresponds to the progressive negative values of $\beta$ that determine a saturation of the slopes permitted in the other FP projections.} When the ETGs become passive quenched objects, with a luminosity decreasing at nearly constant $\sigma$, the galaxies can only move in one direction, { the one determined by the value of $\beta$}.
\end{enumerate}

{ We conclude by saying that with our empirical approach, we are able to explain the problem of the FP tilt and the properties  of all its projections only by invoking an obvious argument: the temporal evolution of galaxies cannot be discarded when we look at the FP and its projections. These diagrams are sensitive to the temporal evolution of galaxies, simply because each individual object can move in a different way according to the value of $\beta$}.

The \Lsigb\ relation is an empirical way to catch such temporal evolution. The values of $\beta$ are related to the history of mass assembly and to the luminosity evolution. {
In the hierarchical context in fact, the role of mass should not be confused with that of luminosity. Mass and luminosity in each galaxy can vary in different ways. For example, in absence of merger events, as time goes by all galaxies progressively decrease in their luminosity while keeping constant their mass}.

In the hierarchical scenario the tilt of the FP and the observed distributions of ETGs in its projections depend on the evolutionary path followed by galaxies of different masses. Only when galaxies become passive and quenched objects they start to approach the distribution expected from the VT. 

{Last but not least we mention here the result obtained by \citet{Donofrioetal2021}, who demonstrated that it is impossible to reproduce the \IeRe\ plane starting from the \Lsiga\ relation and using the values of $L_0$ and $\alpha$ derived from the fit of the FJ. The only way to achieve this is to allow for  much larger variations of such parameters, \ie\ to use the \Lsigb\ relation. For this reason this approach also solves the old problem of connecting the \Lsig\ and \IeRe\ planes}. 

\begin{acknowledgements}
      The authors thank the Department of Physics and Astronomy of the Padua University for the economical support.
\end{acknowledgements}

%
   \bibliographystyle{aa} 
   \bibliography{FP.bib} 
%

\end{document}